\documentclass[pra,showpacs,twocolumn,superscriptaddress,nofootinbib]{revtex4}

\usepackage{amsmath,amsfonts,amssymb,amstext}
\usepackage{graphicx}
\usepackage[colorlinks,linkcolor=blue]{hyperref}
\usepackage{multirow}

\newcommand{\ket}[1]{\left|#1\right\rangle}
\newcommand{\bra}[1]{\left\langle#1\right|}

\newcommand{\avg}[1]{\left\langle #1 \right\rangle}

\def\cO{{\cal O}}
\DeclareMathOperator{\Tr}{Tr}

\def\({\left(}
\def\){\right)}

\def\su{\mathfrak{su}}

\newcommand{\be}{\begin{equation}}
\newcommand{\ee}{\end{equation}}
\newcommand{\bea}{\begin{eqnarray}}
\newcommand{\eea}{\end{eqnarray}}
\newcommand{\bF}{\begin{figure}}
\newcommand{\eF}{\end{figure}}

\newcommand{\bi}{\begin{itemize}}
\newcommand{\ei}{\end{itemize}}
\newcommand{\ud}{\mathrm{d}}
\newcommand{\mbf}[1]{\mathbf{#1}}

\begin{document}
\title{Random unitary maps for quantum state reconstruction}
\date{December 9, 2009}

\author{Seth T. Merkel}
\affiliation{Institute for Quantum Computing, Waterloo, ON N2L 3G1, Canada}

\author{Carlos A. Riofr\'{i}o}
\affiliation{Center for Quantum Information and Control (CQuIC), Department of Physics and Astronomy, University of New Mexico, Albuquerque, NM, 87131, USA}

\author{Steven T. Flammia}
\affiliation{Perimeter Institute for Theoretical Physics, Waterloo, Ontario N2L 2Y5, Canada}

\author{Ivan H. Deutsch}
\affiliation{Center for Quantum Information and Control (CQuIC), Department of Physics and Astronomy, University of New Mexico, Albuquerque, NM, 87131, USA}

\begin{abstract}
We study the possibility of performing quantum state reconstruction from a measurement record that is obtained as a sequence of expectation values of a Hermitian operator evolving under repeated application of a single random unitary map, $U_0$.  We show that while this single-parameter orbit in operator space is not informationally complete, it can be used to yield surprisingly high-fidelity reconstruction.  For a $d$-dimensional Hilbert space with the initial observable in $\mathfrak{su}(d)$, the measurement record lacks information about a matrix subspace of dimension $\ge d-2$ out of the total dimension $d^2-1$.  We determine the conditions on $U_0$ such that the bound is saturated, and show they are achieved by almost all pseudorandom unitary matrices.  When we further impose the constraint that the physical density matrix must be positive, we obtain even higher fidelity than that predicted from the missing subspace.  With prior knowledge that the state is pure, the reconstruction will be perfect (in the limit of vanishing noise) and for arbitrary mixed states, the fidelity is over 0.96, even for small $d$, and reaching $\mathcal{F}>0.99$ for $d>9$.  We also study the implementation of this protocol based on the relationship between random matrices and quantum chaos.  We show that the Floquet operator of the quantum kicked top provides a means of generating the required type of measurement record, with implications on the relationship between quantum chaos and information gain.
\end{abstract}

\maketitle

\section{Introduction}

Quantum state reconstruction (QSR, or quantum tomography) is a fundamental tool in quantum information science that has been carried out in a variety of systems and in a variety of protocols \cite{QSR04, DAriano03}.  The essential procedure of QSR is to use the statistics of the measurement results on an ensemble of identical systems to make a best estimate of the prepared state $\rho_0$.  This can be achieved, e.g., through a series of strong projective measurements of a set of Hermitian observables \cite{Klose2001} or through continuous weak measurement of a time-series of observables \cite{silberfarb05, smith06}.  High fidelity QSR typically requires an ``informationally complete" measurement record.  One can obtain informational completeness by measuring the expectation values of a set of Hermitian operators that span an operator basis for $\rho_0$, or some more general operator ``frame" \cite{DAriano01,roy07}.  Restricting our attention to a Hilbert space of finite dimension $d$, and fixing the normalization of $\rho_0$, the set of Hermitian operators must form a basis for the Lie algebra $\su(d)$.  Laboratory realization of such a record is intimately tied to {\em controllability}, i.e., the ability to reconfigure the apparatus in such a way as to generate arbitrary unitary maps.  In the continuous measurement context, when the system is controllable it is possible to choose control fields for the system such that, when viewed in the Heisenberg picture, the observables evolve over the span of the algebra.  While the necessity of information completeness is rigorous if one requires high fidelity for the reconstruction of all arbitrary states, in a number of situations this condition can be substantially relaxed.  Examples include schemes that are designed to achieve high performance of the reconstruction on average \cite{Aaronson2006}, or over only some restricted set of the state space \cite{gross2009}.  For these protocols, the performance of a restricted set of measurements is often nearly as good as for an informationally complete set of measurements, and yet require dramatically fewer measurement resources. 

In this paper, we study another example of informationally incomplete measurements that nonetheless can be used in a high-fidelity QSR --- measurement of a time-series of operators generated by a single-parameter random evolution.  Consider continuous weak measurement of an observable, $\cO$, through a meter that couples to an ensemble of $N$ identical systems.  The members of the ensemble undergo identical, separable time evolution in a well chosen manner.  Assuming the subsytems remain in a product state, or equivalently that a mean-field approximation is valid, we can write our measurement record quite generally as
\be
\mathcal{M}(t) = N \avg{ \cO}(t) + \delta M(t),
\label{eq:measurement}
\ee 
where,  $\delta M(t)$ describes the deviation from the mean value arising from state-dependent quantum uncertainty and noise in the detection system.   For unitary evolution, $\avg{\cO}(t)= \Tr\left(U^{\dagger}(t) \cO U(t) \rho_0\right)$. The QSR problem is to retrodict the initial state of the system $\rho_0$ from the signal $\mathcal{M}(t)$.   We can simplify the analysis of this problem by considering a discrete set of measurements at intervals $\Delta t$, $\left\{ \cO_n \equiv  U^{\dagger}(n \Delta t) \cO U(n\Delta t)  \right\}$. The ultimate fidelity of the QSR will be limited by the finite signal-to-noise ratio.  While the choice of unitary evolution necessary to determine an arbitrary $\rho_0$ from $\mathcal{M}_n$ is not unique, a necessary and sufficient condition is that the set $\left\{ \cO_n \right\}$  be informationally complete.  A good strategy is to choose the dynamics such that for each $n$, $U(n\Delta t)$ is a random matrix, chosen from an appropriate Haar measure.  In that case our measurement record is not only provably informationally complete but is also unbiased over time.  Suppose, however, we choose $U(n\Delta t)=\left(U_0\right)^n$, where $U_0$ is a {\em fixed} unitary matrix.   In this case, the observable series $\cO_n$ traces out a single orbit in operator space; we call this a one-parameter measurement record.  As we will show, the record is not informationally complete, but nevertheless can lead to high fidelity QSR for all states but a set of small measure if $U_0$ is a random unitary, especially for large dimensional spaces.  These results elucidate the connection between random evolution and information gain at the quantum level.

The remainder of the article is organized as follows.  In Sec. II we show that the measurement operators generated from a single parameter trajectory cannot span the entirety of the operator algebra, $\su(d)$, but that the operators that lie outside of the subspace of the measurement record are a vanishingly small fraction in the limit $d \rightarrow \infty$.  Next, we study the performance of QSR using the weak measurement protocol \cite{silberfarb05} for these incomplete measurement records.  We show that even at small $d$, when one includes the physical constraint of {\em positivity} of the density matrix, QSR performs surprising well for almost all quantum states, well beyond that expected if one had solely considered the vector space geometry of the Lie algebra.  Finally, in Sec. III, we connect these abstract results to physical realizations using the unitary Floquet maps of the quantum kicked top \cite{haake} whose associated classical dynamics is chaotic.  As quantum chaos is associated with pseudorandom matrix statistics, this protocol provides intriguing new signatures of quantum chaos in QSR.

\section{One-Parameter Measurement Records}\label{sec:span}

In this section, we study whether or not information completeness for QSR is achievable from a one-parameter measurement record.  The one-parameter orbit in operator space is defined by the time-series $\cO_n = (U_0^{\dag})^n \cO (U_0^{\phantom \dag})^n $, where $\cO$  is a Hermitian operator and  $U_0$ is a fixed unitary matrix.  We will restrict the observable $\cO$ to have zero trace since the component proportional to the identity gives no useful information in QSR.  We thus ask, is it possible to reconstruct a generic quantum state $\rho_0$ if one can measure the expectation values of all of the observables in the time series?  To answer this, we consider $\mathcal{A}\equiv\text{span}\left\{ \cO_n \right \}$, and determine the size  the orthocomplement subspace with respect to the trace inner product,  $\mathcal{A}_\perp$; operators in this set are not measured in the time-series. Such missing information renders the measurement incomplete, and thus incompatible with perfect QSR, no matter what signal-to-noise ratio is available in the laboratory.

To find the dimension of $\mathcal{A}$, consider the subspace of operators that are preserved under conjugation by $U_0$, $\mathcal{G} \equiv \left\{ g \in \su(d) \left|\, U_0^{\phantom \dag} g U_0^{\dagger} = g \right\}\right.$. Let $\mathcal{B} =\left\{ g \in \mathcal{G} \left|\, \Tr(g \cO) = 0\right\}\right.$.  It thus follows that $\mathcal{B} \subseteq \mathcal{A}_\perp$ since  $\forall g \in \mathcal{B}$  
\be
\Tr(\cO_n g) = \Tr \left( (U_0^{\dagger})^n \cO(U_0^{\phantom \dag})^n g \right) = \Tr(\cO g)  =0.
\ee
As the two spaces are orthogonal, $\dim\mathcal{A} + \dim \mathcal{B} \le \dim (\su(d)) =d^2-1$.  Now,  if $U_0$ has nondegenerate eigenvalues, $\mathcal{G}$ will be isomorphic to the the largest commuting subalgebra of $\su(d)$ (the Cartan subalgebra),  but for degenerate $U_0$, $\mathcal{G}$ will contain additional elements.  Since the Cartan subalgebra has dimension $d-1$, $\dim \mathcal{G} \geq d-1$.  By definition, $\mathcal{B}$ is obtained from $\mathcal{G}$ by projecting out one direction in operator space, and thus  $\dim \mathcal{B} =\dim \mathcal{G}-1 \ge  d-2$.  It follows that 
\be
\dim \mathcal{A} \le \dim (\su(d))-\dim \mathcal{B} \le d^2-d +1.  
\label{eq:bound}
\ee

This is the first principal result -- a one-parameter measurement record is not informationally complete since $\dim \mathcal{A}_\perp>0$ (when $d > 2$).  But, it remains to be seen how much the missing information impacts the fidelity of QSR.  An immediate question is to determine the conditions on $U_0$ and $\cO$ required to saturate bound in Eq.~(\ref{eq:bound}).    Since $U_0$ is a unitary matrix it is always diagonalizable as
\be
 U_0 = \sum_{j=1}^d e^{-i \phi_j} \ket{j}\!\bra{j},
\ee 
and in this basis
$\cO_n$ has the representation 
\be
 \cO_n = \sum_{j,k=1}^d e^{-i n (\phi_j -\phi_k)} \bra{k} \cO \ket{j} \ket{k}\!\bra{j}. 
\ee
The diagonal component has no $n$-dependence, so it is useful to rewrite $\cO_n$ as 
\be 
\cO_n =\sum_{j=1}^d \bra{j} \cO \ket{j} \ket{j}\! \bra{j}   + \sum_{j \neq k }^d e^{-i n(\phi_j - \phi_k)} \bra{k} \cO \ket{j} \ket{k}\!
\bra{j}.\label{eq:On} 
\ee
To show that that $\mathcal{A}$ is spanned by $d^2-d+1$ linearly independent matrices we must have that
\be
\sum_{n=0}^{d^2-d} a_n \cO_n = 0 \qquad \textrm{iff} \qquad a_n = 0 ~~ \forall n.
\ee
We can write this condition out explicitly using Eq. (\ref{eq:On}) as
\bea
&&\left( \sum_{n=0}^{d^2-d} a_n \right)  \sum_{j=1}^d \bra{j} \cO \ket{j} \ket{j}\!\bra{j}   +\nonumber\\
 &&\sum_{j \neq k}^d \left( \sum_{n=0}^{d^2-d} a_n e^{-i n (\phi_j - \phi_k)} \right) \bra{k} \cO \ket{j} \ket{k}\!\bra{j} = 0.
\eea
The system of equations is underconstrained if either $\bra{j} \cO \ket{j}  =0$ for all $j$ or $\bra{j} \cO \ket{k}  =0$ for any $j \neq k$.  Assuming this is not the case, the condition for linear dependence is given by a set of linear equations on $a_n$ of the form
\be\label{eq:systemeq}
\underbrace{\left(\begin{array}{ccccc}
1 & x_0^{\phantom 2} & x_0^2 &\cdots & x_0^{d^2-d}\\
1 & x_1^{\phantom 2} & x_1^2 &\cdots & x_1^{d^2-d}\\
\vdots & \vdots &\vdots &\ddots& \vdots\\
1 & x_{d^2-d}^{\phantom 2} & x_{d^2-d}^2 &\cdots & x_{d^2-d}^{d^2-d}
\end{array}\right) }_V
\left( \begin{array}{c}
a_0\\
a_1\\
\hdots \\
a_{d^2-d}
\end{array} \right) =  0.
\ee
Here we have written $x_0 =1$ and $x_m = e^{-i (\phi_j -\phi_k)}$, for some indexing of the pairs $(j,k)$ to $1\leq m \leq d^2-d$.  
The condition for linear independence is simply $\det V \neq 0$.  Expressed as above, one can see that $V$ is an instance of a Vandermonde matrix, whose determinant is easy to evaluate through the formula \cite{hornandjohnson} 
\be
\det V = \prod_{0 \leq j < k \leq d^2 - d} (x_k - x_j).
\ee 
For our system of equations to become linearly dependent before saturating the previous bound, we would need that $e^{-i (\phi_j - \phi_k)} =
e^{-i (\phi_{j'} - \phi_{k'})}$ for some distinct pair of the couples $(j,k)$ and $(j',k')$, or $e^{-i (\phi_j - \phi_k)} = 1$ for some $(j,k)$. 

In summary, in order for the dimension of the span of a one-parmeter measurement record to saturate the bound of $\dim \mathcal{A} = d^2 -d +1$,  the eigenphases, $\phi_j$, and the eigenvectors, $\ket{j}$, of $U$ must satisfy the following constraints:
\bea
\label{eq:conditions}
	1.&& \exists \ j \ \textrm{s.t.}, \quad \bra{j} \cO \ket{j} \neq 0 \nonumber\\
	2.&& \forall \ j \neq k, \quad \bra{k} \cO \ket{j} \neq 0 \nonumber\\
	3.&& \forall \ j \neq j', \quad \phi_j - \phi_{k} \neq \phi_{j'} - \phi_{k'}  
\eea
Note that the third condition enforces that both the eigenphases, as well as their pairwise differences, must be distinct.  There is an interesting interpretation of these conditions from the perspective of universal control.  While $U_0$ defines a one-parameter trajectory, the set $\{ U_0, e^{i \mathcal{O}} \}$ defines a universal set of unitary matrices that can generate arbitrary maps, if the conditions in Eq.~(\ref{eq:conditions}) are satisfied \cite{altafini2002}.  

If $U_0$ is a unitary matrix chosen randomly from the Haar measure on ${\sf SU}(d)$, the saturation conditions will almost surely be satisfied, independent of $\cO$.  Therefore, a generic unitary evolution will almost always generate a measurement record that spans the full $d^2-d+1$ operators.  In fact, the typical members of many types of pseudorandom ensembles of unitary matrices satisfy these
constraints, e.g. unitaries drawn from $t$-designs or approximate $t$-designs~\cite{Ambainis2007, Gross2007, Dankert2009, Roy2009}, random quantum circuits~\cite{Harrow2009, Brown2009}, as well as unitary evolutions
that possess globally chaotic dynamics in the classical limit \cite{Emerson03, Scott03, haake}.  These types of pseudorandom evolutions are more
readily available in practical situations, providing possible avenues to test these results in laboratory implementations.

The results of this section show that a one-parameter evolution generates a measurement record that misses a subspace of dimension $d-2$ out of the full $\su(d)$ algebra whose dimension is $d^2-1$.  For very large Hilbert space dimensions, the implication is that all but a vanishing fraction of the information regarding measurements on the quantum system is contained in this type of record.  The reconstruction fidelity is a fairly complicated nonlinear function of the measurement record,
\be
	\mathcal{F}(\rho_{\rm est}, \rho_0) = \bigg[\Tr \bigg(\sqrt{\sqrt{\rho_{\rm est}} \rho_0 \sqrt{\rho_{\rm est}}}\bigg)\bigg]^2.
\ee
It is not clear that this fidelity will be directly related to the fraction of operator space spanned by our set of observables.    Surprisingly, the situation is in fact more favorable than this na\"{\i}ve assumption.  In the next section we will see that merely requiring the reconstructed density matrix be positive provides a powerful constraint, allowing us to use a one-parameter measurement record induced by the orbit of a single pseudorandom unitary matrix to perform very high fidelity reconstructions even for small dimensional Hilbert spaces, for all but a very small subset of states.

\section{Density Matrix Reconstruction from an Incomplete Measurement}\label{S:Reconstruct}
The QSR protocol we consider was first proposed by Silberfarb {\em et al.} \cite{silberfarb05} and implemented by Smith {\em et al.} \cite{smith06}.  In this scenario one has access to an ensemble of  $N$ identically prepared systems all initialized to the same state, $\rho_0$.  The system is weakly measured yielding the record given in Eq.~(\ref{eq:measurement}).  For sufficiently weak coupling, the deviation of the measurement result from the quantum expectation value is dominated by the noise on the detector (e.g., shot noise of a laser probe) rather than the quantum fluctuations of measurement outcomes intrinsic to the state (known as projection noise).  In this case, there is negligible backaction on the quantum state during the course of the measurement, and the ensemble remains factorized.  We treat the detector noise as Gaussian white noise defined as $\delta M(t)=\sigma W(t)$, where $W(t)$ is a Weiner process and $\sigma$ determines the noise variance.  

We examine a stroboscopic time-series of the measurement record, where the observables evolve according to the one-parameter trajectory discussed in the previous section.  At discrete times $t=n \Delta t$,
\be
M_n=N \Tr (\cO_n \rho_0) + \sigma W_n,
\label{eq:record}
\ee
where $\cO_n$ and $\rho_0$ are Heisenberg operators.  The problem is thus reduced to one of stochastic estimation of $\rho_0$ given $\left\{ M_n \right\}$.  To accomplish this, the density operator is expanded in a basis of Hermitian matrices, $\left\{ E_\alpha \right\}$, so $\rho_0=\sum_\alpha r_\alpha E_\alpha +I/d$.  We take $\rho_0$ to be unit trace, explicitly removing the identity from the operator basis.  Defining a rectangular matrix with elements $\tilde{\cO}_{n\alpha}=\Tr (\cO_n E_\alpha)$, the measurement time-series given in Eq.~(\ref{eq:record}) can be expressed as the vector $\mbf{M} =N \tilde{\cO} \mbf{r} + \sigma \mbf{W}$.  Because the fluctuations around the mean are Gaussian distributed, the maximum-likelihood-estimate of the unknown parameters $\left\{ r_\alpha \right\}$ is the least-squared fit, given by the vector
\be
\mbf{r}^{\rm ML} = \frac{1}{N}  (\tilde{\cO}^T \tilde{\cO} )^{-1} \tilde{\cO}^T \mbf{M}.\label{eq:fit}
\ee
If the measurement record is informationally complete, which occurs when the covariance matrix $\mathcal{C}=\tilde{\cO}^T \tilde{\cO}$ has full rank $d^2-1$, and in the absence of measurement noise, the maximum-likely-estimate, $\rho^{\rm ML}=\sum_\alpha r_\alpha^{\rm ML} E_\alpha+I/d$, is exactly $\rho_0$.  If the set of observables $\left\{\cO_n  \right\}$ is not informationally complete, then $\mathcal{C}$ is not full rank and we must replace the inverse in Eq.~(\ref{eq:fit}) with the Moore-Penrose pseudo-inverse, i.e., inverting only over the space in which the covariance matrix has support.  In this case, a state with support on the null space of $\mathcal{C}$ will yield a maximum likelihood estimate with sub-unit reconstruction fidelity.   

In the presence of measurement noise, or when the measurement record is incomplete, the estimate $\rho^{\rm ML}$ can have negative eigenvalues which are unphysical.  To obtain a better reconstruction of the state, we seek the physical density matrix ``closest"  to $\rho^{\rm ML}$.  We use the covariance matrix $\mathcal{C}$ as a cost function or metric to measure the distance between  $\rho^{\rm ML}$ and new estimate $\bar{\rho}$ by defining
\be \label{E:costnorm}
	\| \mathbf{r}^{\rm ML}-\bar{\mathbf{r}}\|^2 = (\mathbf{r}^{\rm ML}-\bar{\mathbf{r}})^T \mathcal{C}(\mathbf{r}^{\rm ML}-\bar{\mathbf{r}}) .
\ee
Technically speaking, this quantity is not a norm but rather a seminorm, meaning that there exist some vectors $\mathbf{v}$ such that $\| \mathbf{v} \| = 0$ but $\mathbf{v} \not=0$.  This cost penalizes us for taking displacements in directions that increase the variance in the measurement uncertainty.  We thus need an $\bar{\mbf{r}}$ that minimizes the distance $\| \mathbf{r}^{\rm ML}-\bar{\mathbf{r}}\|$ while respecting the constraint $\sum_{\alpha} \bar{r}_\alpha E_\alpha + I/d \geq 0$.  While there is generally no analytic solution to this problem, the optimization is a semidefinite program which is efficiently solvable numerically \cite{vandenberghe96, sedumi}. 

\begin{figure}[t]
\begin{center}
\includegraphics[width=8cm,clip]{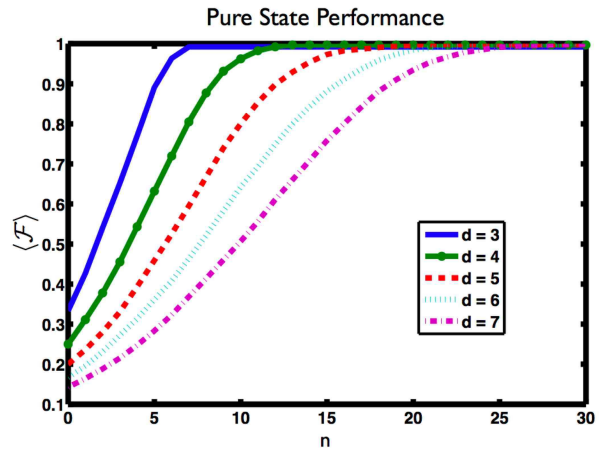}
\caption{(Color online) Numerical simulations the QSR protocol for pure states as function of $n^{th}$ expectation value measured in the time-series, and for different dimensions of the Hilbert space, $d$. Each data point represents the average reconstruction fidelity of 100 pure states drawn from the Fubini-Study measure, additionally averaged over measurement records derived from ten different Harr-random unitary propagators. }
\label{F:pure}
\end{center}
\end{figure}

The condition of positivity is a powerful constraint that describes correlations between observables that can lie along orthogonal directions in operator space.  For example, in the case of a 2-level quantum system, if $\avg{\sigma_z} = 1$, positivity implies $\avg{\sigma_x} = \avg{\sigma_y}=0$, fully specifying the state from a single expectation value.  In the context of noisy measuremets, the positivity constraint allows us to perform high fidelity QSR in the face of uncertainty by enforcing consistency conditions on our measurement outcomes.  When we consider incomplete measurement records, positivity can place bounds on the means of observables which otherwise would be completely undetermined.  This can greatly increase the fidelity of QSR.  Intuitively, while many vectors $\bar{\mathbf{r}}$ might minimize Eq.~(\ref{E:costnorm}), only very few of these are also compatible with positivity.  As we show below, in the context of a one-parameter measurement record generated by a single random matrix, the requirement of positivity provides substantial additional information leading to very high fidelity of QSR, well beyond what one would na\"{\i}vely predict.

We are now prepared to quantitatively analyze the performance of our QSR protocol in the case of the one-parameter measurement record arising from an incomplete set of observables that satisfy Eq.~(\ref{eq:conditions}).  As our system, we consider a system with spin $J$  described by a Hilbert space of dimension $d=2J+1$.  We will fix $\cO = J_z$ and select a random unitary matrix $U_0$ from the Haar measure on ${\sf SU}(d)$.   Such a random matrix will almost always satisfy the constraints of Eq.~(\ref{eq:conditions}), except for a set of measure zero.  As our goal is to determine how the information missing in a subspace of observables impacts the QSR fidelity, we will simplify the analysis by assuming that noise on the measurement  is vanishingly small.  We study the performance of different classes of states, randomly chosen by an appropriate measure.  For each set of states, we will look at the average fidelity between the initial and reconstructed states, $\langle \mathcal{F} \rangle = \int \ud \rho_0 \mathcal{F}(\bar{\rho}, \rho_0)$, where $\ud \rho_0$ is a measure on the space of density operators.

The simplest case to analyze is when we have prior information that $\rho_0$ is a pure state, $\ket{\psi_0}$.  Figure~\ref{F:pure} shows the average fidelity as one sequentially measures the expectation value of the $n^{\textrm{th}}$ observable in the series, for different dimensions of the Hilbert space $d$.  Averages are taken for 10 choices of random unitary matrices, each of which is averaged over 100 random pure states distributed on the Fubini-Study measure \cite{zyczkowski2006}.  Two striking features are seen in these plots:  (i) unit fidelity is achieved for any $d$ even though the record was said to be informationally incomplete; (ii) the protocol reconstructs the state well before we measure all $d^2-d+1$ independent observables. The inclusion of positivity dramatically improves the reconstruction fidelity for pure states.  In fact, a one-parameter measurement record generated by a random $U_0$ can be used to reconstruct almost all pure states perfectly in the absence of noise.  

The performance of the QSR protocol can be understood given the prior information we have assumed.  A pure state is specified by $2d-2$ real parameters, whereas we measure $d^2-d+1$ expectation values.  Thus, it should come as no surprise that the measurement record contains enough information to reconstruct the state.   In fact, in this case one can use positivity to explicitly recover the missing information exactly from the measurement record, without resorting to the numerical semidefinite program discussed above.  In general, the missing information is associated with matrices that commute with $U_0$.  Thus, when expressed in the eigenbasis of $U_0$, only the diagonal matrix elements of the density operator might not be estimated.  A necessary (but not generally sufficient) condition for a matrix $\rho_0$ to be positive semidefinite is that its matrix elements must satisfy the following set of inequalities: $\rho_{ii} \rho_{jj} - |\rho_{ij}|^2 \geq 0 $, i.e.~all of the $2 \times 2$ matrix minors must be positive semidefinite \cite{hornandjohnson}.  If additionally the state is pure,  these inequalities become equalities.  Therefore, if any of the off-diagonal matrix elements are nonzero, we can completely determine all of the diagonal elements via the equations  $\rho_{ii} = \left(|\rho_{ij}| |\rho_{ik}|\right)/|\rho_{jk}|$.  A special case is if all of the off-diagonal matrix elements of $\rho_0$ are zero. Then the state must be one of the eigenvectors of $U_0$, but such states lie in a set of measure zero.  Measurements that are informationally complete solely for pure states are called PSI-compete~\cite{Flammia2005, Finkelstein2004}.  

\begin{figure}[t]
\begin{center}
\includegraphics[width=8cm,clip]{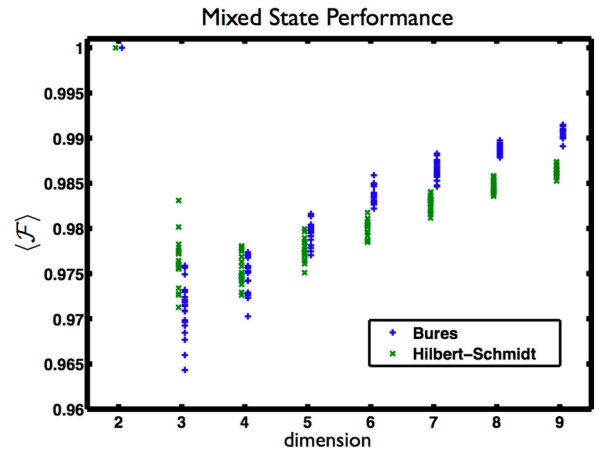}
\caption{(Color online) Fidelity of the QSR protocol for mixed states as a function of the dimension of Hilbert space $d$.  Each point represents the average reconstruction fidelity from a measurement record of length $10(d^2-d+1)$ (a long-time limit) generated from a different random unitary propagator from the Harr measure.   We average over 200 density matrices drawn from the Bures (blue crosses) or Hilbert-Schmidt (green x's) measures.  For each dimension we show the average fidelities from twenty of these measurement procedures.  }
\label{F:Bures}
\end{center}
\end{figure}

While one can easily explain the high-fidelity performance of the QSR protocol in the case of a pure state, for mixed states, this is far from clear, and the power of the positivity constraint comes fully to the fore.  For mixed states, and $d>2$, the average fidelity of our reconstruction will never reach unity because some density matrices cannot be reconstructed from the information in our incomplete measurement record.  For example, some convex combinations of eigenstates of $U_0$ are indistinguishable from the maximally mixed state, even though the fidelity between the two can be very small.  Nonetheless, as we see below, the one-parameter measurement record generated by a single random unitary still performs very well on average, even for generic mixed states.

Figure~\ref{F:Bures} shows the average fidelity for two choices of measures on density matrices, the Bures measure and the Hilbert-Schmidt measure \cite{braunstein94, zyczkowski2006}, with states sampled according to the construction provided by Osipov, Sommers and \.{Z}yczkowski \cite{Osipov09}.  For both distributions we look at a long-time limit of the time-series, here $10 (d^2-d+1)$ measurement steps, and plot the average fidelity as a function of the dimension of the Hilbert space rather than $n$.  In the limiting case of negligible noise on the measurement, we have already extracted all the possible information about the state after $d^2-d+1$ measurements.  In practice, increasing the measurement record serves to smear out the information over the measured observables, leading to a more uniform distribution for the non-zero eigenvalues of $\mathcal{C}$, which is numerically favorable. As seen in these two plots, on average, the one-parameter measurement records perform surprisingly well.  In all cases the mean fidelity is greater than $0.96$ with a minimum around $d=3$ or $d=4$.  After this dip, the minimum of the fidelity looks to be monotonically increasing with the size of the Hilbert space.  Additionally, the particular instantiation of the random unitary map appears to make very little difference (less than 0.01 fidelity), with the residual difference decreasing as the dimension increases.   

\section{Example: Quantum Kicked-Top}

In Sec.~\ref{sec:span} we discussed that the conditions given in Eq.~(\ref{eq:conditions}) can be satisfied by a pseudorandom unitary matrix, instead of a true random matrix sampled from the Haar measure.  One such class of matrices are the Floquet maps associated with ``quantum chaos", i.e., periodic maps whose classical dynamical description shows a globally chaotic phase space.  An example is the quantum kicked-top  \cite{haake}, a system that recently has been realized in a cold atomic ensemble \cite{Chaudhury2009}.  In this section we explore how our QSR protocol performs in this context, providing a possible route to laboratory studies, and novel signatures of chaos in quantum information.

The standard quantum kicked top (QKT) dynamics consists of a constant quadratic twisting of a spin (``top"), punctuated by a periodic train of delta-kicks of the spin around an orthogonal axis.  The Floquet operator for this perodic map is typically written as the product of noncommuting unitary matrices,
\be
U_{\rm QKT} = e^{- i \phi J_z^2 / J} e^{-i \theta J_x}.\label{eq:floquet}
\ee
The parameters $\theta$ and $\phi$ represent the angles of linear and nonlinear rotation respectively.  The dynamics exhibit a classically chaotic phase space for an appropriate choice of these parameters  \cite{haake}.  The connection between chaos in this system and random matrices has been well studied, particularly, the relationship between the level statistics of the Floquet eigenvalues, chaos, and symmetry.   Floquet maps associated with global chaos are random matrices that divide into different classes.  If the Floquet operator is time-reversal invariant, the level statistics are that of the circular orthogonal ensemble (COE); without additional symmetry they are members of the circular unitary ensemble (CUE). The latter group is ${\sf U}(d)$ or ${\sf SU}(d)$ depending on the context.  The measurement records generated from matrices chosen from either the COE or CUE will satisfy the eigenvalue conditions described in Eq.~(\ref{eq:conditions}) almost surely.  The QKT is known to have a time-reversal symmetry and classically chaotic dynamics. It still does not have COE statistics in the full $d=2J+1$ Hilbert space, however, due to an additional symmetry. The QKT map is invariant under a $\pi$-rotation about the $x$-axis, leading to a parity symmetry.    This system therefore has a doubly degenerate eigenspectrum, breaking the conditions in Eq.~(\ref{eq:conditions}).  While such Floquet operators generate a measurement record that have much less information relative to an arbitrary state, we can perform high-fidelity QSR in for states restricted to a subspace defined by the additional symmetry, here, the states that have even parity under reflection around the $x$-axis.  To do this we require that our initial operator $\cO$ is also symmetric under reflection, e.g., $\cO = J_x$.

\begin{figure}[t]
\begin{center}
\includegraphics[width=8cm,clip]{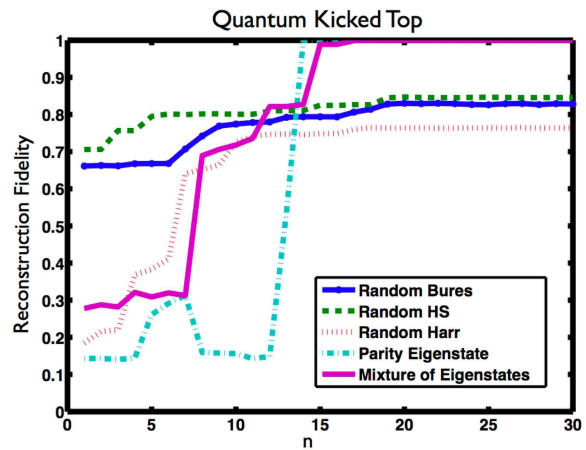}
\caption{(Color online) Reconstruction fidelity for a variety of states versus $n$.  The initial measurement observable is $J_x$, and each subsequent observable whose expectation value we measure is obtained by evolving under the Floquet map of a quantum kicked top, Eq.~(\ref{eq:floquet}).  For generic random states in the whole Hilbert space (pure or mixed), the reconstruction preforms poorly.  Density matrices that are invariant with respect to $\pi$-rotation around the x-axis, such as the cat state, an eigenstate of the parity operator, or an incoherent mixture of odd-parity $J_x$ eigenstates, are reconstructed with
high fidelity. }
\label{F:qkt}
\end{center}
\end{figure}

\begin{figure}[t]
\begin{center}
\includegraphics[width=8cm,clip]{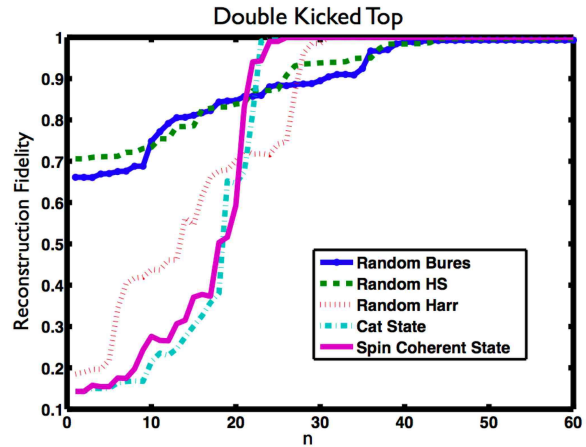}
\caption{(Color online) Same as Fig.~\ref{F:qkt} when the evolution is given by the double kicked top.  All states, pure or mixed, asymptote to fidelities near unity, with the pure states reaching their maxima quicker than the mixed states.  }
\label{F:dkt}
\end{center}
\end{figure}

We present examples of this type of reconstruction in Fig.~\ref{F:qkt}.  Here we look at the QKT dynamics for a spin $J=3$ particle  (a $d=7$ dimensional Hilbert space).  We choose the parameters $\phi = 7$ and $\theta = 0.228$, values for which the classical phase space is known to be globally chaotic.  Additionally, we let the noise on the measurement approach zero. For general mixed and pure states sampled from the full Hilbert space, this reconstruction performs poorly.  However, if we restrict our attention to states that are eigenstates of parity, we reconstruct with near unit fidelity.  As a check, we find that covariance matrix $\mathcal{C}$ has rank 19.  This agrees with our predictions, since this space has a 4-fold degenerate $-1$ parity-eigenspace and a 3-fold degenerate $+1$ parity-eigenspace.  The measurement operators that preserve this symmetry must be block diagonal, lacking the $(3 \times 4) $ components from each of the two off-diagonal blocks.  For this dimension  $d^2 - d+1 - 24 = 19$.    
  
We can examine the effects of a pseudorandom unitary that satisfies Eq.~(\ref{eq:conditions}) on the whole space if we look at the ``double kicked top" where we alternate kicking about $x$ and $y$.  Here the Floquet operator has the form
\be
U_{\rm 2KT} = e^{- i \phi J_z^2 / J } e^{-i \theta_x J_x}e^{- i \phi' J_z^2 /J} e^{-i \theta_y J_y}.\label{eq:double}
\ee
If we choose $\phi = \phi'= 6$, $\theta_x = \pi /2$ and $\theta_y = 0.228$, these operators have no time-reversal symmetry, or indeed any other symmetry, and so approximate the spectrum of the CUE.  Since the double kicked top Floquet operator shares no symmetries with $J_z$ we can choose $\cO = J_z$ in order to satisfy the first condition of Eq.~(\ref{eq:conditions}).

In Fig.~\ref{F:dkt} we show the QSR performance for a time-series generated by the double kicked top in a Hilbert space of a spin-3 particle.  Here the fidelity asymptotes to unity for all of our choices of  states,  as we would expect from the simulations in Sec.~\ref{S:Reconstruct}.  The reconstruction reaches its asymptotic fidelity when we have made $d^2-d+1 = 43$ measurements.  As we saw previously, the pure states require less measurements to reach their asymptotic value and the QSR can be perfect in the absence of noise.

\section{Summary and outlook}

We have studied measurement records that are derived by stroboscopically measuring the expectation values of a single observable of a system that is evolving under the repeated application of a single unitary map.  We have shown that this record never contains complete information about the quantum state, however for unitary maps chosen randomly or pseudorandomly, only a vanishing fraction of the information is missing.  When combined with the constraint of positivity, this incomplete measurement record led to a protocol for quantum state reconstruction that had high-fidelity performance for typical mixed and pure quantum states.  For pure states we can achieve unit fidelity reconstruction (in the absence of noise) and for mixed states the fidelity is greater than 0.99 for $d>9$.

 A particular set of pseudorandom matrices we studied in some detail are the Floquet maps generated by the quantum delta-kicked top.  In the general case, these maps appear to be equally as effective for reconstruction as Harr-random unitary maps.  In cases where the kicked top map exhibited additional symmetry, we were able to see that our reconstruction protocol required the extra constraint that the measurement operator and states shared the symmetry as well.  A map that is chaotic on the whole phase space saturates the bound of pseudo-random unitary operators on the whole Hilbert space.

It is surprising that such a simple measurement protocol should lead to such  good average reconstruction fidelities.  The reason for this appears to be a combination of the mixing power of random evolutions and the constraints on state space associated with positivity.  We do not yet have a rigorous explanation of these results, however, because the set of positive operators is a convex cone rather than a vector subspace~\cite{zyczkowski2006}, and thus it is difficult to quantify the volume of states that have both large support in the missing subspace of $\su(d)$ and are positive.  Our conjecture is that both the increasing average fidelity and the decreasing dependence on the sampled unitary can be a explained based on concentration of measure in analogy with the work in~\cite{gross2009}.  Essentially, most randomly sampled states have very little support on a the subspace that we do not measure.  Related  work in the field of matrix completion has shown that such ``incoherence'' between states and (incomplete) measurements can provably lead to high-fidelity state reconstructions especially when the states in question are low rank~\cite{Gross2009b} or near to low-rank states~\cite{gross2009}.  Irrespective of a rigorous proof, it is an empirical fact that our protocol works well with typical states and typical unitary maps.  We expect that as the dimension of the Hilbert space increases, almost all of the states and unitary evolutions that are sampled will be very close to a typical value, resulting in high QSR fidelity.

\acknowledgments 
We thank Andrew Scott for helpful discussions.  STM was supported by Industry Canada and QuantumWorks.  STF was supported by the Perimeter Institute for Theoretical Physics.  Research at Perimeter is supported by the Government of Canada 
through Industry Canada and by the Province of Ontario through the Ministry of Research~\& Innovation.  STF thanks the Kavli Institute for Theoretical Physics where this work was completed; this research is also supported in part by the National Science Foundation under Grant No. PHY05-51164.  IHD and CAR where support by the National Science Foundation, Grant 0903692.


\begin{thebibliography}{10}
\bibitem{sedumi}
We used the the Matlab programs YALMIP and SeDuMi for solving the semidefinite
  programs.
\end{thebibliography}

\end{document}